\pdfoutput=1

\documentclass[aps,showpacs,preprintnumbers,amsmath,amssymb,
eqsecnum,
twocolumn, tightenlines
]{revtex4}

\usepackage{graphicx}

\newcommand{\be}{\begin{eqnarray}}
\newcommand{\ee}{\end{eqnarray}}

 \newcommand{\gsim}{\mathrel{\hbox{\rlap{\lower.55ex \hbox {$\sim$}}
                   \kern-.3em \raise.4ex \hbox{$>$}}}}
\newcommand{\lsim}{\mathrel{\hbox{\rlap{\lower.55ex \hbox {$\sim$}}
                   \kern-.3em \raise.4ex \hbox{$<$}}}}

\def\roughly#1{\mathrel{\raise.3ex\hbox{$#1$\kern-.75em%
\lower1ex\hbox{$\sim$}}}}
\def\lsim{\roughly<}
\def\gsim{\roughly>}

\begin{document}


\title{ The Lifetime of the Electric Flux Tubes \\ near the QCD Phase Transition }
\author {Cyrus Faroughy and Edward Shuryak}
\address {Department of Physics and Astronomy, State University of New York,
Stony Brook, NY 11794}
\date{\today}

\begin{abstract}
Electric flux tubes are a well known attribute of the QCD vacuum in which they manifest confinement of electric color charges. Recently, experimental results have appeared suggesting that not only those objects persist at temperatures $T\approx T_c$ near the QCD phase transitions, but their decay is suppressed and the resulting clusters in AuAu collisions are larger than in pp (i.e. in vacuum). This correlates well with recent theoretical scenarios that view the QCD matter in the $T\approx T_{c}$ region as a dual-magnetic plasma dominated by color-magnetic monopoles. In this view the flux tubes are stabilized by dual-magnetic currents and are described by dual-magnetohydrodynamics (DMHD). In this paper we calculate classically the  dissipative effects in the flux tube. Such effects are  associated with rescattering and finite conductivity of the matter. We derive the DMHD solution in the presence of dissipation and then estimate the lifetime of the electric flux tubes. The conclusion of this study is that a classical treatment leads to too short of a lifetime for the flux tubes.
\end{abstract}
\maketitle

\section{Introduction}
\subsection{Motivation}
Non-perturbative phenomena in both the QCD vacuum and in the finite temperature/density QCD matter have been the subject of intense studies for a long time. Various phenomenological approaches have been proposed during the last few decades to tackle them.  One simple example is the stochastic QCD vacuum model   \cite{Simonov} which provided a  picture of the gluonic "condensate" and correlations as a constant fields, another is the instanton liquid model \cite{Schafer} that explained several phenomena related to the $SU(N_{F})$ and the $U_{A}(1)$ symmetry breaking.

Most researchers in the field  hold to a generic picture of confinement introduced by t'Hooft and Mandelstam in the 1980's: the so-called ``dual superconductivity". It suggests that confinement is a dual Meissner effect, and that a ``coil" which prevents the color-electric flux tube from spreading out is produced by a magnetic supercurrent.
The condensate of Cooper pairs produced in usual superconductors is substituted by a Bose-Einstein condensate of some objects possessing color magnetic charge.
 Presently, however, there is no concrete understanding of the field configurations that are accountable for color confinement. The effort for identifying the corresponding topological objects responsible for color confinement is still continuing to this date. Monopoles, dyons, or their composites, induced by fermions, have been shown to provide a mechanism for confinement and chiral symmetry breaking in the $\cal N$=2 SYM theory \cite{Seiberg:1994aj}. But it is still unclear whether such objects can be concretely carried into the QCD-like theories without involving scalars.
The seminal paper by Polyakov \cite{Polyakov} set an early example of confinement in (2+1)D Yang-Mills theories by means of gluomagnetic monopoles. Recently, Unsal \cite{Unsal:2010zz} has proposed an interesting (3+1)D extension of the latter model of confinement for QCD-like theories, involving fermions, by invoking certain composite objects endowed with total magnetic and zero topological charges. However, here too it remains to be seen whether such objects can be truly responsible for confinement, particularly in the framework of lattice gauge theories.

In the last few years, this subject has been revitalized by experimental studies at the Relativistic Heavy Ion Collider (RHIC), where hot QCD matter is produced and studied. One conclusion stemming from
the analysis of RHIC results is that the quark-gluon plasma (QGP) in the $1$-$2T_{c}$ temperature interval is a strongly coupled fluid (sQGP).
Liao and Shuryak \cite{Liao_ES1} (a review on the subject can also be found in \cite{Shuryak:2008eq}) have related this finding with the so called ``magnetic scenario", arguing that while
color-electric objects (quarks and gluons) interact stronger and stronger when approaching the critical temperature $T_{c}$, the Dirac condition implies that the color-magnetic objects should become lighter and more  weakly coupled. They conjectured that electric components dominate the matter at high-$T$ while the magnetic components dominate it near $T_c$. Furthermore, they gave arguments that an equilibrium point in which both electric and magnetic coupling constants are equal ($\alpha_{m}=\alpha_{e}=1)$ exists at $T\approx  1.4T_{c}$. Lattice observations have confirmed that monopoles form a relatively strongly coupled liquid where the magnetic coupling $increases$ at high $T$ (see discussion in \cite{Liao_ES2}). Lattice observations of monopoles also strongly support the idea that they form a Bose-Einstein condensate at exactly $T=T_c$ \cite{D'Alessandro}.

   Recent two and three particle correlations in experiments at RHIC indicate that certain fluctuations occurring on top of the overall
expanding quark-gluon plasma (QGP) have small or even zero expansion velocity in the near-$T_{c}$ region, suggesting the existence of stabilized electric flux tubes. In addition, the experimental results also show that the resulting clusters in AuAu collisions are even larger than in pp collisions (i.e. in vacuum). The purpose of this paper is to use the classical dual-magnetohydrodynamic model to study their evolution when the dissipative effects are included. We will also attempt to evaluate the stability of the flux tubes in the near-$T_{c}$ region, where we consider the plasma to be dual-magnetized \cite{Liao_ES1}. Specifically, uncondensed color-magnetic monopoles circulate and form solenoidal currents which, by dual-Faraday's law, induce a color-electric field and create the flux tube. By studying the diffusion of this field in the medium we can calculate the halflife of the flux tube and see if the field is strong enough to account for its stability during this magnetized phase. The dissipative effects are included by allowing the conductivity of the matter to be finite. 

\subsection{Dual-Magnetohydrodynamics}
Magnetohydrodynamics (MHD) studies the dynamics of electrically conducting fluids and is widely used in plasma physics. In other words, it studies the interaction between a magnetic field and a plasma, treating it as a continuous medium. It is an approximation which keeps only the magnetic field in Maxwell's equations, while the electric field is entirely screened. Furthermore, MHD is used by solar physicists to describe the overall effects of electric currents and magnetic fields in the Sun's corona to describe the  sun spots and flux tubes.

As mentioned previously, one scenario has been proposed where the QCD matter near the $T_{c}$ region is dual-magnetized. In this region, the electric and magnetic screening masses are assumed to be $M_m > M_e$
because the color-magnetic monopoles dominate the scene. 
(This is opposite to the case of high temperatures, in which electric particles --
quarks and gluons  -- dominate, and therefore $M_m \sim  g^{2}T \ll M_e\sim gT$, where $g$ is the gauge coupling, small at high $T$. The high-T value for $M_m$ kas been suggested by Polyakov \cite{Polyakov}) 
   It therefore is plausible to use DMHD ignoring the magnetic fields. 

In the usual plasmas like in the Sun, the electric screening mass is very high and thus electric fields are ignored and magnetic fields are included.
Under certain conditions the flux tubes can be formed as a result of solenoidal electric currents. 
Analogously,  in  RHIC collisions, there exist a  "dual corona" 
  \cite{Shuryak:2009cy} in which the electric flux tubes are produced by the magnetic currents. 
  In what follows we study the diffusion of the field for flux tubes in MHD and then translate the results in the context of DMHD to investigate the lifetime of the flux tubes in the near-$T_{c}$ region of the QCD matter.

\section{The Diffusion equation with finite Conductivity}
The ideal MHD approximation is the limit of  $infinite$ conductivity of the plasma $\sigma\rightarrow\infty$, analogous to the $zero$ viscosity approximation in the case of ideal hydrodynamics. In this section, for self-consistency of the paper, we include the known derivation of the diffusion equation for the magnetic field in standard MHD with \textit{finite} conductivity in order to account for dissipative effects. The fundamental equations of MHD are:
\begin{equation}
-\frac{\partial \vec{B}}{\partial t} = \vec{\nabla} \times \vec{E}
\end{equation}
\begin{equation}
E = -\vec{v} \times \vec{B} + \frac{\vec{j}}{\sigma}
\end{equation}
and
\begin{equation}
\vec{j} = \vec{\nabla} \times \frac{\vec{B}}{4 \pi}
\end{equation}
If the currents are orthogonal to the magnetic field (as they are in an ideal solenoid), then $\vec{v} \times \vec{B} =0$ and the above equations simplify into the Diffusion Equation:
\begin{equation}
\frac{\partial \vec{B}}{\partial t} = \eta\nabla^{2}\vec{B}
\end{equation}
 where we have used $div \vec B=0$ and
 $\eta = 1/4 \pi \sigma$ is the magnetic diffusivity. The magnetic diffusivity contains the conductivity of the plasma which will be discussed in part IV. The dual version of the diffusion equation is the same as the latter with the
magnetic field $\vec{B}$ replaced by the dual-electric field $\tilde E$. In addition, instead of having ions and electrons with different masses
($m_{i}>>m_{e}$), there are two (uncondensed) color-magnetic monopoles and antimonopoles with equal mass and opposite charge $g_{+}$ and $g_{-}$
forming two currents of equal magnitude circulating along the cylinder in opposite directions.

\section{Solving the diffusion equation}
The flux tube is stabilized by magnetic currents circulating solenoidally on its surface. We place the solenoid in a cylindrical coordinate system with its vertical axis along the z direction. At t=0 each current $\pm\vec{J}$ flows circularly in the $\pm\vec\varphi$ direction at $r=a_{o}$, the initial radius of the tube. The initial condition is taken to be a gaussian profile of the field centered at the origin.
At all times, the field points only along the $z$ axis. We assume azimuthal symmetry so that there is no $\varphi$ dependence and consider the field to be constant along the $z$ direction. Therefore the field $\tilde E(r,t)$ will only depend on the radial direction $r$ and time $t$.
The diffusion equation in cylindrical coordinates is:

\begin{equation}
\frac{1}{r}\frac{\partial}{\partial r}(r\frac{\partial \tilde E(r,t)}{\partial r}) = \frac{1}{\eta}\frac{\partial \tilde E(r,t)}{\partial t}
\end{equation}
where
\begin{equation}
\eta = \frac{1}{4 \pi \sigma}
\end{equation}
is now the dual diffusivity and $\sigma$ is the  dual conductivity of the plasma. The initial profile of the field is assumed to be a Gaussian centered at the origin:
\begin{equation}
\tilde E(r,0)=\tilde E_{o}e^{-\frac{r^{2}}{a_{o}^{2}}}
\end{equation}
Separation of variables $\tilde E(r,t)=R(r)T(t)$ decouples the PDE $(3.1)$ into two ordinary ODE's, one with time dependence and one with spatial dependence:
$$\frac{\partial T(t)}{\partial t} + \lambda^{2}\eta T(t)=0$$
$$\frac{\partial}{\partial r}(r\frac{\partial R(r)}{\partial r}) + \lambda^{2}rR(r)=0$$
The first equation gives us the time dependent part $T(t)=e^{-\lambda^{2}\eta t}$ and the second equation is the Bessel equation of zeroth order ($m=0$) and the solution is a superposition of Bessel functions of the first and second kind. We exclude the Bessel function of second kind since they are not finite at $r=0$ and obtain $R(t)=AJ_{0}(\lambda r)$. In Sturm-Liouville conditions the general solution is:
\begin{equation}
\tilde E(r,t)= \int_{0}^{\infty}A(\lambda)J_{0}(\lambda r)e^{-\lambda^{2}\eta t}d\lambda
\end{equation}
To find the $A(\lambda)$'s lets look at $\tilde E(r,t)$ at $t=0$:
$$\tilde E(r,0)= \int_{0}^{\infty}A(\lambda)J_{0}(\lambda r)d\lambda$$
Now multiply both sides by $J_{0}(\beta r)r$ and integrate over $r$:

$$\int_{0}^{\infty}\tilde E(r,0)J_{0}(\beta r)rdr = \int_{0}^{\infty}A(\lambda)\int_{0}^{\infty}J_{0}(\beta r)J_{0}(\lambda r)rdrd\lambda$$

The \textit{closure relation} for Bessel functions gives:

$$\int_{0}^{\infty}J_{0}(\beta r)J_{0}(\lambda r)rdr = \frac{1}{\beta}\delta(\beta-\lambda)$$

and we obtain:

$$\int_{0}^{\infty}\tilde E(r,0)J_{0}(\beta r)rdr = \int_{0}^{\infty}A(\lambda)\frac{1}{\beta}\delta(\beta-\lambda)d\lambda = \frac{1}{\beta}A(\beta)$$

$$A(\lambda) = \lambda\int_{0}^{\infty}\tilde E(r,0)J_{0}(\lambda r)rdr$$
which for $\tilde E(r,0)=\tilde E_{o}e^{-\frac{r^{2}}{a_{o}^{2}}}$ gives:

$$A(\lambda) = \lambda \tilde E_{o}\int_{0}^{\infty}e^{-\frac{r^{2}}{a_{o}^{2}}}J_{0}(\lambda r)rdr = \frac{1}{2}\,\tilde E_{{0}}\lambda\,{a_{o}}^{2}{{\rm e}^{-\frac{1}{4}\,{\lambda}^{2}{a_{o}}^{2}}}$$

Plugging this back in $(3.4)$ we obtain an expression for $\tilde E(r,t)$:

$$\tilde E(r,t)= \int_{0}^{\infty}\frac{1}{2}\,\tilde E_{0}\lambda\,{a_{o}}^{2}{{\rm e}^{-\frac{1}{4}\,{\lambda}^{2}{a_{o}}^{2}}}J_{0}(\lambda r)rdrJ_{0}(\lambda r)e^{-\lambda^{2}\eta t}d\lambda$$

and finally:
\begin{equation}
\tilde E(r,t)= \frac{\tilde E_{o}a_{o}^{2}}{{a_{o}}^{2}+4\eta t}exp{\left[-\frac{r^{2}}{a_{o}^{2}+4\eta t}\right]}
\end{equation}
Comparing $(3.3)$ with $(3.5)$, for $t>0$, the term $a_{o}^{2}$ acquires an effective time dependent form $a_{o}^{2} + 4\eta t$ that shows an increase in the field's squared radius. Therefore the field's time dependent radius is given by $a(t) = (a_{o}^{2} + 4\eta t)^{1/2}$.

\section{Estimates of the conductivity and the fate of the flux tubes}
The standard electrodynamic plasma consists of electrons and ions having densities $N_e$ and $N_i$ respectively. Due to large difference in masses,
$m_e\ll m_i$ the current is assumed to be due to the motion of electrons only.
The textbook expression for the conductivity of the usual plasmas, normal to the magnetic field \cite{LPtextbook} is:
\be \sigma_\perp= {3\sqrt{\pi} e^2 N_e \over \sqrt{2} m \nu_e }\ee
for $Z=1$ ions, and
\be \nu_e={4\pi  e^4 L_{e} N_i\over m^{1/2} T^{3/2} } \ee
is the collision rate between electron and ions.  $L_{e}$ is the Coulomb logarithm and it is equal to  $\ln(1/\chi_{min})$ where $\chi_{min}$ is the magnitude of the smallest  angles for which the scattering can still be regarded as Coulomb scattering. The electron-electron collisions are ignored since they cannot change the total momentum of the electrons and thus do not modify the electron current.
Our task is to translate these results into those appropriate for the QCD matter in the near-$T_c$ region.

It consists of four main components. The first two are positively and negatively charged monopoles. Their densities are denoted by $N_+$ and $N_-$. As stressed before, the monopoles create counterdirected flows around the electric flux tube.
The medium produced has overall zero magnetic charge, so that $N_+=N_-=N_m/2$.
The last two components are electric objects -- quarks and gluons -- with densities $N_q$ and $N_g$ respectively. As noted in the Introduction,
the electric components dominates at high $T$, in the Quark-Gluon Plasma, but not in the  near-$T_c$ region, where they are suppressed.
The total scattering rate for the positive monopoles is
\be \nu_+= \nu_{+-} + \nu_{+g} + \nu_{+q} \ee
and, by symmetry, the total collision rate is then $\nu_{tot}=\nu_{+}$.

It is important to note that the $++$ scattering term is omitted for the same reason as the $ee$ collisions are omitted in the electrodynamic plasmas: they would not change the total current.
The $+-$ cross section is the transport cross section due to magnetic Coulomb forces, so the $+-$ collision rate is given by dual to the $electron-ion$ rate above.
It is obtained by the substitution of the coupling $e^2\rightarrow g_m^2/4\pi$ (note the difference in $4\pi$ resulting from two
different ways the fields are defined in QED and QCD) and the density $N_e\rightarrow N_-$.

Scattering on electric objects is different, as discussed in details by Ratti and Shuryak \cite{Ratti:2008jz}. It has similar Rutherford-like scattering at small angles but the transport cross section is dominated by large (near-backward) scattering angles. In order to obtain expressions for the scattering rates and conductivity, we have to use certain empirical values of the parameters involved. All of them are, in principle, a function of the temperature $T$, but we restrict our discussion to the vicinity of $T\approx T_c$. The values for the monopole density, magnetic Coulomb coupling and the mass that we use for the estimates are as follows:\\

\begin{tabular}{|l|c|r|}\hline
quantity & value & reference \\ \hline
${N_m/T_c^3}$ & $\approx 1$ & Fig.1 of Ref.\protect\cite{Ratti:2008jz} \\\hline
 ${g_m^{2}/4\pi}$  & $\approx 4/5 $ &     Fig.3  of  Ref.\protect\cite{Liao_ES2} \\\hline
${m/T_c} $  & $\approx 2$   &  Fig.6 of Ref.\protect\cite{D'Alessandro}\\ \hline
$\nu_{+g}/T_{c}$ & $\approx 2$ &  Fig.14 of Ref.\protect\cite{Ratti:2008jz}\\ \hline
\end{tabular}
\vskip .3cm

In addition, we assume that $\nu_{+g}\approx\nu_{+q}$ \footnote{(22) Although the monopole-quark scattering rate has not yet been calculated, we may take it to be of the order of the monopole-gluon scattering rate.}. From these relations we obtain the +- dual Coulomb collision rate:
$$ \nu_{+-}={g_{m}^4 L_{e} N_-\over 4\pi m^{1/2} T_{c}^{3/2}} = {g_{m}^4 L_{e} \sqrt{2}N_-\over 4\pi 2T_{c}^{2}}= {g_{m}^4L_{e}T_{c}\sqrt{2}\over 16\pi}$$
And from equation ($4.3$):
$$ \nu_{tot}=  {g_{m}^4L_{e}T_{c}\sqrt{2}\over 16\pi} + 2T_{c}  + 2T_{c}=T_{c} \left(\frac{g_{m}^{4}L_{e}\sqrt{2} + 64\pi}{16\pi }\right)$$

The conductivity $(4.1)$ is then:
$$ \sigma= {3\sqrt{\pi} g_{m}^2 N_- \over 4\pi \sqrt{2} m \nu_{tot} }={3\sqrt{2\pi} g_{m}^2 T^{2}_{c} \over 32\pi \nu_{tot} }= {3\sqrt{2\pi} g_{m}^2 T_{c} \over 2(g_{m}^{4}L_{e}\sqrt{2} +64\pi)}$$

The value we used for $T_c$ is the $T_c=170$ MeV from regular QCD with quarks rather than the larger $T_c=260$ MeV value from pure gauge.
Using $g_m^{2}/4\pi \approx 4/5$ and estimating $\chi_{min} = 1/10 $ so that $L_{e}= \ln(10) \approx 2.30 $, we get:
$$ \nu_{+-}={16\pi L_{e}T_c\sqrt{2} \over 25} \approx 1113.04 MeV = 5.65 fm^{-1}$$
$$ \nu_{tot}= T_c(\frac{16\pi L_e\sqrt{2} + 100}{25}) \approx 1793.04 MeV= 9.10 fm^{-1}$$
$$ \sigma= {3\sqrt{2\pi} T_c \over  \frac{32\pi}{5}(L_e\sqrt{2} + \frac{25}{4\pi})}\approx 12.13 MeV= 6.16 \times 10^{-2} fm^{-1}$$
Finally, from equation $(3.2)$:

$$\eta = \frac{1}{4\pi\sigma} \approx 6.56 \times 10^{-3} MeV^{-1}=1.29 fm $$

Now we need to address the overall timing of the heavy ion collisions at RHIC. According to (very successful) hydrodynamical
simulations, the duration of the magnetized phase of the collisions is $\tau_M\approx 4-5\, fm/c$, at any centrality and any position of the fluid cell. According to our solution the mean square radius of the tube during this time is increasing by
\be a_o^2 \rightarrow a_o^2+4\eta \tau_M \ee
The energy per unit length $\epsilon$ diffused during the magnetic phase is given by
$$ \epsilon(t)=2\pi\int_{0}^{\infty}\left[\frac{1}{8\pi}\,\tilde E(r,t)^{2} + K(r,t)\right] rdr$$

Where $K = (1/2)\rho v^{2}\propto  J(r,t)^{2}$ is the kinetic energy density associated to the monopole current $\vec{J}(r,t)$. Using equation (2.3) we see that the time component of $K$ scales like the time component of the magnetic energy. As a result, during the magnetic phase the total energy per unit length scales as :
$$ \frac{\epsilon(\tau_{M})}{\epsilon(0)}= {\frac {{a_{o}}^{2}}{ {a_{o}}^{2}+4\,\eta\,\tau_{M}  }} $$

We estimate $a_o\approx 0.5 fm$ so that $a_o^2 \approx 0.25 fm^{2}$ and from the values above we get:
$$4\eta \tau_M \approx 23.26 fm^{2}$$
To gain more insight about the lifetime of the flux tube, let us calculate the half-life $t_{1/2}$ of the field at the origin. We know that at the origin and at $t=0$ we have $\tilde E(0,0)=\tilde E_{o}$. So, by definition:
\begin{equation}
\tilde E(0,t_{1/2})= \frac{\tilde E_{o}}{2} = \frac{\tilde E_{o}a_{o}^{2}}{{a_{o}}^{2}+4\eta t_{1/2}}
\end{equation}
and solving for $t_{1/2}$:
\be t_{1/2} = \frac{1}{4} \frac{a_{o}^{2}}{\eta} = \pi \sigma a_{o}^{2} = 5 \times 10^{-2}fm \ee
which is many times shorter than the expected lifetime of the observed flux tubes.

\section{Summary and Discussion}
In this paper we have used a $classical$ dual-magnetohydrodynamic approach to calculate the flux tube lifetime in the magnetic phase of the QGP near the QCD phase transition.
More specifically, we have found a solution for the flux tube including the dissipative ``diffusive" term. We calculated the value for the
``dual magnetic diffusion constant" using a picture of monopole-monopole and monopole-gluon rescattering and found that this crude classical rescattering model predicts very strong dissipative effects that are
way too strong for the flux tubes to survive in the few fm/c timeframe of the magnetic phase. Yet, ``ridge" correlations of the detected pions are found in experiments.

One possible view on this, held e.g. by the BNL group  \cite{Dumitru:2008wn,Gavin:2008ev}, is that the flux tubes do indeed decay very quickly, as the estimates above suggest, and the observed ``ridge" is nothing but the extra amount of entropy left behind. A problem with this interpretation (discussed by one of us in \cite{Shuryak:2009cy}) is that a spot of extra thermal entropy/energy would evolve hydrodynamically into a cylinder of several fm radius, which is in direct contradiction with the rather narrow $\phi$ distribution width of the ridge.

More likely, a classical approach to the flux tube dissipation is incorrect. First of all, unlike flux tubes usually considered by magnetohydrodynamics (e.g. in solar plasmas) the QCD flux tubes under consideration are small in size and not larger than the quasiparticle Compton wave length. This suggests that one should use a quantum-mechanical description, like the one in \cite{Liao:2007mj}.
Quantum effects in the monopole motion may provide two ``supercurrents" which propagate through each other without $any$ dissipation.
We know that this is the case  in the confining vacuum.

 Finally, let us also mention that there are additional evidence for survival of the flux tubes 
  in the magnetic phase, at   $T\approx T_c$, which do not come  from RHIC experiments but from lattice numerical simulations of the QCD thermodynamics.
 Those indicate that  baryonic states remain under such conditions, and that their density of state is well described by Hagedorn-like ``stringy" states, see  \cite{Lin:2009ds}.
Unfortunately we do not know  the lifetime of these baryons  as the lattice thermodynamics does not give us such information.

    \section*{Acknowledgments}
      The work of ES is supported by the US DOE grant DE-FG-88ER40388.

\end{document}